\documentclass[11pt]{article}

\usepackage[paper=a4paper,dvips,top=2.0cm,left=2.2cm,right=2.2cm,bottom=2.9cm,footskip=1.4cm,voffset=1.2cm]{geometry}
\usepackage{amsfonts,amsmath,amsthm,amssymb}
\numberwithin{equation}{section}  
\usepackage[svgnames]{xcolor}
\usepackage{fix-cm}
\usepackage{sectsty}
\usepackage{fancyhdr}
\usepackage{lastpage}
\usepackage{graphicx}
\usepackage{url}
\usepackage{enumerate}
\usepackage{paralist}
\usepackage{multicol}
\usepackage{color}  
\usepackage{float}  
\usepackage{epsfig}
\usepackage{hyperref}   
\hypersetup{colorlinks=true,linkcolor=beamer@PRD, citecolor=beamer@PRD}
\usepackage{authblk}  
\usepackage{cite}  

\newcommand\myref[1]{\textcolor{beamer@PRD}{(}\ref{#1}\textcolor{beamer@PRD}{)}}

\definecolor{beamer@blue}{RGB}{0,0,255}
\definecolor{beamer@mediumblue}{RGB}{0,0,190}
\definecolor{beamer@midnightblue}{RGB}{25,25,112}
\definecolor{beamer@navy}{RGB}{0,0,128}
\definecolor{beamer@darkblue}{RGB}{0,0,139}
\definecolor{beamer@purple}{RGB}{128,0,128}
\definecolor{beamer@levander}{RGB}{100.,149.,237.}
\definecolor{beamer@PRD}{RGB}{46,48,146}
\definecolor{beamer@green}{RGB}{0,128,0}
\definecolor{beamer@darkgreen}{RGB}{0,100,0}
\definecolor{beamer@olive}{RGB}{128,128,0}
\definecolor{beamer@darkolivegreen}{RGB}{85,107,47}
\definecolor{beamer@gray}{RGB}{190,190,190}
\definecolor{beamer@ivry}{RGB}{220,220,220}
\definecolor{beamer@new}{RGB}{40,120,50}
\definecolor{shadecolor}{RGB}{220,220,220}
\definecolor{beamer@darkslategray}{RGB}{47,79,79}
\definecolor{beamer@chocolate}{RGB}{210,105,30}
\definecolor{beamer@brown}{RGB}{165,42,42}
\definecolor{beamer@orangered}{RGB}{255,69,0}
\definecolor{beamer@maroon}{RGB}{128,0,0}
\definecolor{beamer@white}{RGB}{234,242,243}
\definecolor{beamer@silver}{RGB}{0.5,0.45,0.37}

\begin{document}

\title{\textbf{Squeezed atom laser for Bose-Einstein condensate with minimal length}}
\author{\textbf{Sanjib Dey$^\circ$ and V\'eronique Hussin$^\bullet$} \\ \small{$^\circ$Department of Physics, Indian Institute of Science Education and Research Mohali, \\ Sector 81, SAS Nagar, Manauli 140306, India \\ $^\bullet$D\'epartement de Math\'ematiques et de Statistique $\&$ Centre de Recherches Math\'ematiques, \\ Universit\'e de Montr\'eal, QC, H3C 3J7, Canada \\ E-mail: dey@iisermohali.ac.in, veronique.hussin@umontreal.ca}}
\date{}
\maketitle
\thispagestyle{fancy}
\begin{abstract}
We study a protocol for constructing a squeezed atom laser for a model originating from the generalized uncertainty principle. We show that the squeezing effects arising from such systems do not require any squeezed light as an input, but the squeezing appears automatically because of the structure of the model it owns. The output atom laser beam becomes squeezed due to the nonlinear interaction between the Bose-Einstein condensate and the deformed radiation field created due to the noncommutative structure. We analyze several standard squeezing techniques based on the analytical expressions followed by a numerical analysis for further insights.    
\end{abstract}	 
\section{Introduction} \label{sec1}
\addtolength{\footskip}{-0.2cm} 
After many decades since its inception, quantum theory yet serves as an endless source of contemporary phenomena in the physical world as well as it keeps enriching in its mathematical structure. With the advent of laser and quantum optics, the theory has become more popular, since it gives rise to the opportunity of testing many fundamental properties of quantum mechanics through simple experiments  \cite{Aspect_Dalibard_Roger,Ou_Pereira_Kimble_Peng,Furusawa_et_al}. A crucial prerequisite for many of these tests is the ability to create squeezed states, as they are the only source of continuous variable quantum entanglement \cite{Ou_Pereira_Kimble_Peng,Bowen_Lam_Ralph}. There exist several methods of generating squeezed states by using the standard techniques of quantum optics, which do assist the procedure of verifying the fundamental properties. However, the atom laser \cite{Ketterle_Noble} provides the freedom to revisit many of these tests using massive particles rather than the photons. For instance, in \cite{Kheruntsyan_Olsen_Drummond}, the authors have shown that the continuous variable entanglement between the amplitude and phase of spatially separated atomic beams for the Einstein-Podolsky-Rosen (EPR) tests can be generated by dissociating a molecular Bose-Einstein condensate (BEC), or by outcoupling from a BEC using a Raman transition with squeezed light \cite{Haine_Hope}. The tests of fundamental quantum theories by using atom laser not only provide an alternative avenue over quantum optics, but also it promises improved accuracy in interferometry \cite{Peters_Chung_Young_Hensley_Chu,Ketterle}, quantum shot-noise limit for the measurement of quadrature squeezing \cite{Dowling}, etc.

The squeezed atom laser is created usually by transferring the squeezed state of light onto the atoms by coupling the atoms out of the BEC and into the atom laser beam \cite{Jing_Chen_Ge,Fleischhauer_Gong,Haine_Olsen_Hope,Tan_Li_Ficek}, and indeed there have been partial success in generating them in recent experiments \cite{Ries_Brezger_Lvovsky,Tanimura_et_al,Hetet_et_al, Mccormick_Boyer_Arimondo_Lett}. But, the drawback of such a scheme is that it is required to construct a squeezed light at a particular frequency at which the atoms make up the BEC, which is certainly challenging. Meanwhile, an alternative scheme has also been suggested by utilizing a nonlinear interaction caused by atom-atom scattering to create a Kerr squeezing effect \cite{Jing_Chen_Ge_1,Johnsson_Haine}, thus, removing a significant source of complexity.

In this article, we propose yet another alternative way to obtain a squeezed atom laser by using a model, which not only has a strong mathematical ground originating from the quantum group, but also it demonstrates ample interesting results in generating squeezed states and in other aspects of quantum optics \cite{Dey_Review,Dey_Fring_squeezed,Dey_Hussin,Dey_Hussin_2,Dey_Fring_Hussin}. Previously, it was shown that the optical squeezing properties are inherited in these models by construction, here we show that it is true for the atom laser case also. The models are motivated by the noncommutative structure of space-time, which were introduced to solve the issue of ultraviolet divergence in quantum field theory \cite{Snyder,Seiberg_Witten}. A much simpler form of these models is familiar as minimal length now a days, which are one of the most promising candidates in explaining the Planck scale phenomena and quantum gravity. 

In what follows, we shall discuss the detailed technicalities for the construction of the atom laser in the given scenario and then analyze its squeezing properties. In Sec. \ref{sec2}, we recall the introductory notions of the origin of the model and detailed mathematical and physical consistencies. Sec. \ref{sec3} consists of the detailed discussion for the construction of atom laser for the models that are introduced in Sec. \ref{sec2}. In Sec. \ref{sec4}, we analyze the quantum statistical properties in order to gain insights on the squeezing properties of the atom laser. Finally, our conclusions are stated in Sec. \ref{sec5}. 
\section{Minimal length and generalised uncertainty principle}\label{sec2}
\lhead{Squeezed atom laser for BEC with minimal length}
\chead{}
\rhead{}
\addtolength{\voffset}{-1.2cm} 
\addtolength{\footskip}{0.2cm} 
The concept of minimal length originally follows from the Snyder's noncommutative version of space-time \cite{Snyder}, which was introduced in the essence of implementing a natural cut-off in the ultraviolet domain to renormalize the theory of quantum fields. Today, the theory is about 70 years old and, it has evolved from time to time to reveal its usefulness in different contexts \cite{Connes_Book,Garay,Seiberg_Witten,Madore_Book,Amelino_Majid, Douglas_Nekrasov,Szabo}. A more recent and simpler version that modifies the Heisenberg uncertainty relation to a generalized version, so-called \textit{generalized uncertainty principle} (GUP) is of great interest now a days. In such cases, the space-time commutation relation often involves higher power of momentum and, thus, leads to the existence of nonzero minimal uncertainties in position coordinate, which is familiar as \textit{minimal length} in the literature \cite{Kempf_Mangano_Mann,Das_Vagenas,Gomes_Kupriyanov,Bagchi_Fring, Quesne_Tkachuk,Pedram_Nozari_Taheri}. These models provide a wide-range of applications in several areas of modern physics, particularly in gravitation and black holes \cite{Mead,Maggiore,Scardigli,Ng_VanDam,Park,Bhat_Dey_Faizal_Hou_Zhao}, cosmology \cite{Rovelli,Dzierzak}, string theory \cite{Veneziano,Gross_Mende,Amati,Yoneya,Konishi}, path integral quantum gravity \cite{Greensite,Padmanabhan}, doubly-special relativity \cite{Amelino,Magueijo}, quantum optics and information theory \cite{Dey_Fring_squeezed,Dey_Hussin,Dey_Fring_Hussin} and many more. Moreover, there have been several experimental proposal to test such theories by using opto-mechanical setup \cite{Pikovski,Dey_NPB,Dey_SciRep}. For further interest on the subject, one may follow some review articles in the context, for instance \cite{Garay,Ng_Review,Hossenfelder_Review}.   

Here we consider one such minimal length model in $1D$ obeying the commutation relation \cite{Kempf_Mangano_Mann,Bagchi_Fring,Dey_Fring_Gouba}.
\begin{equation}\label{NCCom}
[X,P]=i\hbar(1+\check{\tau}P^2), \quad X=(1+\check{\tau}p^2)x, \quad P=p,
\end{equation} 
with $\check{\tau}=\tau/m\omega\hbar$ having the dimension of inverse squared momentum and $\tau$ being dimensionless. The corresponding observables $X$ and $P$ are represented in terms of the usual canonical variables $x,p$ satisfying $[x,p]=i\hbar$. The model \myref{NCCom} has been explored extensively to construct coherent states, squeezed states, etc., as well as their physical implications have been discussed  \cite{Dey_Fring_squeezed,Dey_Hussin,Dey_Fring_Hussin}. One interesting aspect of such model is that it can be associated with standard nonlinear generalized models by using some mathematical techniques. However, unlike the usual nonlinear generalization, there are some interesting consequences that follow from such minimal length model. For instance, it has been shown that the squeezed states constructed out of these models are intrinsically more squeezed than the usual quantum optical models, for further details one may refer to a review article in the context \cite{Dey_Review}.

Nevertheless, in order to apply these models in the construction of atom laser, let us discuss the associated mathematical detail briefly. Let us introduce a set of ladder operators $A,A^\dagger$
\begin{eqnarray} \label{Ladder1}
A &=& af(\hat{n})=f(\hat{n}+1)a, \\
A^\dagger &=& f(\hat{n})a^\dagger=a^\dagger f(\hat{n}+1), \notag
\end{eqnarray}
with $f(\hat{n})$ being an operator valued function of the number operator $\hat{n}=a^\dagger a$ so that their action on the Fock state are given by
\begin{equation} \label{Ladder2}
A|n\rangle=\sqrt{n}f(n)|n-1\rangle, \quad A^\dagger|n\rangle=\sqrt{n+1}f(n+1)|n+1\rangle.
\end{equation}
Thus, if a Hamiltonian can be factorized as $A^\dagger A=a^\dagger a f^2(a^\dagger a)$, the corresponding eigenvalues can be expressed as $e_n=\hbar\omega nf^2(n)$. So far, the ladder operators have been defined generically. We now introduce a specific model, namely the harmonic oscillator in the minimal length scenario
\begin{equation}\label{NHHam}
H=\frac{P^2}{2m}+\frac{m\omega^2}{2}X^2-\hbar\omega\left(\frac{1}{2}+\frac{\tau}{4}\right),
\end{equation}
which satisfies \myref{NCCom}. In order to find the ladder operators of this oscillator, one needs to find the eigenvalues. However, since the Hamiltonian \myref{NHHam} is non-Hermitian, its solution is not so straightforward. Nevertheless, it is possible to obtain real eigenvalues if one applies the standard techniques of $PT$-symmetric non-Hermitian systems \cite{Bender_Boettcher,Bender_Making_Sense} and the notions of pseudo-Hermiticity \cite{Scholtz_Geyer_Hahne,mostafazadeh1}. The idea is to construct an isospectral Hermitian Hamiltonian $h$ corresponding to the non-Hermitian Hamiltonian $H$ given by \myref{NHHam}, by considering the non-Hermitian Hamiltonian $H$ to be pseudo-Hermitian. More precisely, if the Hermitian and the non-Hermitian Hamiltonians are related by the similarity transformation $h=\eta H\eta^{-1}$, and one finds the corresponding metric $\eta$ such that $\eta^\dagger \eta$ is a positive definite operator, then the eigenvalues of $h$ belong to the similar class to those of $H$. In our case, the metric is found to be $\eta=(1+\check{\tau}p^2)^{-1/2}$ so that the corresponding Hermitian Hamiltonian turns out to be
\begin{equation}\label{HerHam}
h=\eta H\eta^{-1}=\frac{p^2}{2m}+\frac{m\omega^2}{2}x^2+\frac{\omega\tau}{4\hbar}(p^2x^2+x^2p^2+2xp^2x)-\hbar\omega\left(\frac{1}{2}+\frac{\tau}{4}\right)+\mathcal{O}(\tau^2).
\end{equation}
The eigenvalues of $h$ \myref{HerHam} and $H$ \myref{NHHam} were computed in \cite{Bagchi_Fring,Dey_Fring_Gouba,Dey_Fring_squeezed} as $E_n=\hbar\omega n(C+Dn)+\mathcal{O}(\tau^2)$ upto the first order of $\tau$ in Rayleigh-Schr\"odinger perturbation theory. For more details in this regard, we refer the readers to \cite{Dey_Review}. Coming back to our discussion, the ladder operators of the harmonic oscillator associated with the minimal length scenario can be defined by \myref{Ladder1} and \myref{Ladder2} for
\begin{equation}\label{fn}
f(n)=\sqrt{C+Dn}, \qquad C=1+D, \quad D=\frac{\tau}{2},
\end{equation}
where $\tau$ is the deformation parameter with the degree of noncommutativity being controlled by it. Having understood the ladder operators of the harmonic oscillator in the minimal length case, we can now set out for the construction of atom laser in the given scenario.
\section{Atom laser in minimal length scenario}\label{sec3}
Before discussing the case corresponding to the minimal length, first let us briefly recall the original linear model for the atom laser. The simplest case is to consider a two-level atom having states, say, $|g\rangle$ and $|e\rangle$, where the initial Bose-Einstein condensation (BEC) occurs in the trapped state $|g\rangle$. Whereas, the state $|e\rangle$ remains unconfined from the magnetic trap, but is coupled to the trapped state $|g\rangle$ by a one-mode squeezed optical field tuned near the $|g\rangle \rightarrow |e\rangle$ transition. Within this setup an experiment was performed at MIT \cite{Mewes}, where a BEC in a superposition of trapped and untrapped hyperfine states by a radio frequency (rf) pulse was created. The corresponding second quantized Hamiltonian for a Bose gas containing $N$ identical two-level atoms in the rotating wave approximation acquires the form \cite{Changpu}
\begin{equation}\label{Ham1}
H=\omega_r a^\dagger a+\omega_0 b_e^\dagger b_e+\Omega\left(ab_e^\dagger b_g+a^\dagger b_g^\dagger b_e\right),
\end{equation}
with $\hbar=1$. Here, $b_g^\dagger (b_g)$ and $b_e^\dagger (b_e)$ are the creation (annihilation) operators of the atoms for the magnetically trapped state $|g\rangle$ and the untrapped state $|e\rangle$, respectively, with the level difference being $\omega_0$. Whereas, $a^\dagger$ and $a$ are the creation and annihilation operators of the rf pulse of frequency $\omega_r$, which couples $b_g^\dagger (b_g)$ with $b_e^\dagger (b_e)$ through the matrix element $\Omega=\sqrt{\omega_r/2\epsilon_0 V}$, where $\epsilon_0$ is the permitivity of the vacuum and $V$ is the effective mode volume. The Hamiltonian \myref{Ham1} can be simplified further if we ignore the slow change of the large number of the initial condensate atoms $N_c$ in the ground state, i.e. if we take the Bogoliubov approximation \cite{Bogoliubov}. Under this approximation, one can replace both of the ladder operators of the ground state $b_g^\dagger, b_g$ by the $c$-number $\sqrt{N_c}$, so that the Hamiltonian \myref{Ham1} is transformed to
\begin{equation}\label{Ham2}
H=\omega \left(a^\dagger a+ b^\dagger b\right)+\Omega\sqrt{N_c}\left(ab^\dagger +a^\dagger b\right),
\end{equation}
where $\omega_r=\omega_0=\omega$ as well as $b_e$ has been redefined to $b$ to simplify our notation. The minimal length scenario can restructure the Hamiltonian further, if we consider that the rf pulse is deformed according to the notions of the minimal length model. This is equivalent to replacing the usual ladder operators $a,a^\dagger$ by the deformed operators $A, A^\dagger$ defined by \myref{Ladder2} along with \myref{fn}. Within this framework, the effective Hamiltonian takes the form
\begin{equation}\label{Ham3}
H_d=\omega \left(A^\dagger A+ b^\dagger b\right)+\Omega\sqrt{N_c}\left(Ab^\dagger +A^\dagger b\right).
\end{equation}
Now, if we replace the generalized ladder operators $A,A^\dagger$ in terms of $a,a^\dagger$ as defined by \myref{Ladder1} followed by a substitution of the function $f(n)$ from \myref{fn}, we obtain
\begin{equation}\label{Ham4}
H=\omega \left[\left(C+Da^\dagger a\right)a^\dagger a+ b^\dagger b\right]+\Omega\sqrt{N_c}\left[ab^\dagger\sqrt{C+Da^\dagger a}+\sqrt{C+Da^\dagger a}~a^\dagger b\right].
\end{equation}
In order to achieve an approximate solution of the Hamiltonian \myref{Ham4}, let us introduce the slowly varying polariton operators $M(t)$ and $N(t)$ \cite{Kuang_Ouyang,Jing_Chen_Ge_1}
\begin{equation}\label{TRAFO1}
M(t)=\frac{1}{\sqrt{2}}e^{-2i\Omega\sqrt{N_c}t}[a(t)+b(t)], \qquad N(t) = \frac{1}{\sqrt{2}}e^{2i\Omega\sqrt{N_c}t}[a(t)-b(t)],
\end{equation}
where the operators $M(t), N(t)$ satisfy the conventional commutation relations $[M,M^\dagger]=[N,N^\dagger]=1$, with the rest of the commutators being vanished. The inverse transformation  of \myref{TRAFO1} reads as
\begin{eqnarray}\label{ab}
a(t) &=& \frac{1}{\sqrt{2}}\left[e^{2i\Omega\sqrt{N_c}t}M(t)+e^{-2i\Omega\sqrt{N_c}t}N(t)\right], \\
b(t) &=& \frac{1}{\sqrt{2}}\left[e^{2i\Omega\sqrt{N_c}t}M(t)-e^{-2i\Omega\sqrt{N_c}t}N(t)\right], \notag
\end{eqnarray}
which when replaced in \myref{Ham4} and the oscillating terms like $M^\dagger N, N^\dagger M,$ etc. are being neglected, we arrive at 
\begin{alignat}{1}\label{Ham5}
H &=\left\{\frac{\omega}{4}(2C+D+2)+\frac{\alpha_1}{2D}(D+4C)\right\}M^\dagger M+\left\{\frac{\omega}{4}(2C+D+2)-\frac{\alpha_1}{2D}(D+4C)\right\}N^\dagger N \notag\\ 
&~~+\frac{D}{4}\left(\omega+\frac{2\alpha_1}{D}\right)(M^\dagger M)^2+\frac{D}{4}\left(\omega-\frac{2\alpha_1}{D}\right)(N^\dagger N)^2+2\alpha_2M^\dagger MN^\dagger N,
\end{alignat}
with $\alpha_1=\Omega D\sqrt{N_c}/(2\sqrt{C}), \alpha_2=\omega D/2$. The solution of the Heisenberg equation of motion for the operators $M(t)$ and $N(t)$
\begin{alignat}{1}
\frac{dM(t)}{dt}=i[H,M(t)], \qquad \frac{dN(t)}{dt}=i[H,N(t)],
\end{alignat}
are given by
\begin{alignat}{1}
M(t)&=M(0)e^{-i \left\{\frac{\omega}{2}(C+D+1)+\frac{2\alpha_1}{D}\left(C+\frac{D}{2}\right)+\frac{D}{2}\left(\omega+\frac{2\alpha_1}{D}M^\dagger M\right)+\omega DN^\dagger N\right\}t}, \label{M1}\\
N(t)&=N(0)e^{-i \left\{\frac{\omega}{2}(C+D+1)-\frac{2\alpha_1}{D}\left(C+\frac{D}{2}\right)+\frac{D}{2}\left(\omega-\frac{2\alpha_1}{D}M^\dagger M\right)+\omega DN^\dagger N\right\}t}. \label{N1}
\end{alignat}
By replacing \myref{M1} and \myref{N1} into \myref{ab}, we obtain the expression of the light field operator as follows
\begin{equation}\label{AtomSol}
a(t)=\frac{1}{\sqrt{2}}\left[M(0)e^{-i\Theta_M t}+N(0)e^{-i\Theta_N t}\right],
\end{equation}
where
\begin{alignat}{1}
\Theta_M &=\frac{\omega}{2}(C+D+1)+\alpha_3+(\alpha_1+\alpha_2)M^\dagger M+2 \alpha_2 N^\dagger N, \\
\Theta_N &=\frac{\omega}{2}(C+D+1)-\alpha_3+(\alpha_2-\alpha_1)M^\dagger M+2 \alpha_2 N^\dagger N,
\end{alignat}
with $\alpha_3=\alpha_1+\Omega\sqrt{N_c}(2+\sqrt{C})$. Therefore, we obtain the atom laser solution of the radiation field given by \myref{AtomSol}, which can be analyzed further to verify its squeezing properties.
\section{Quantum statistical squeezing properties}\label{sec4} 
\begin{figure}
\centering \includegraphics[scale=0.64]{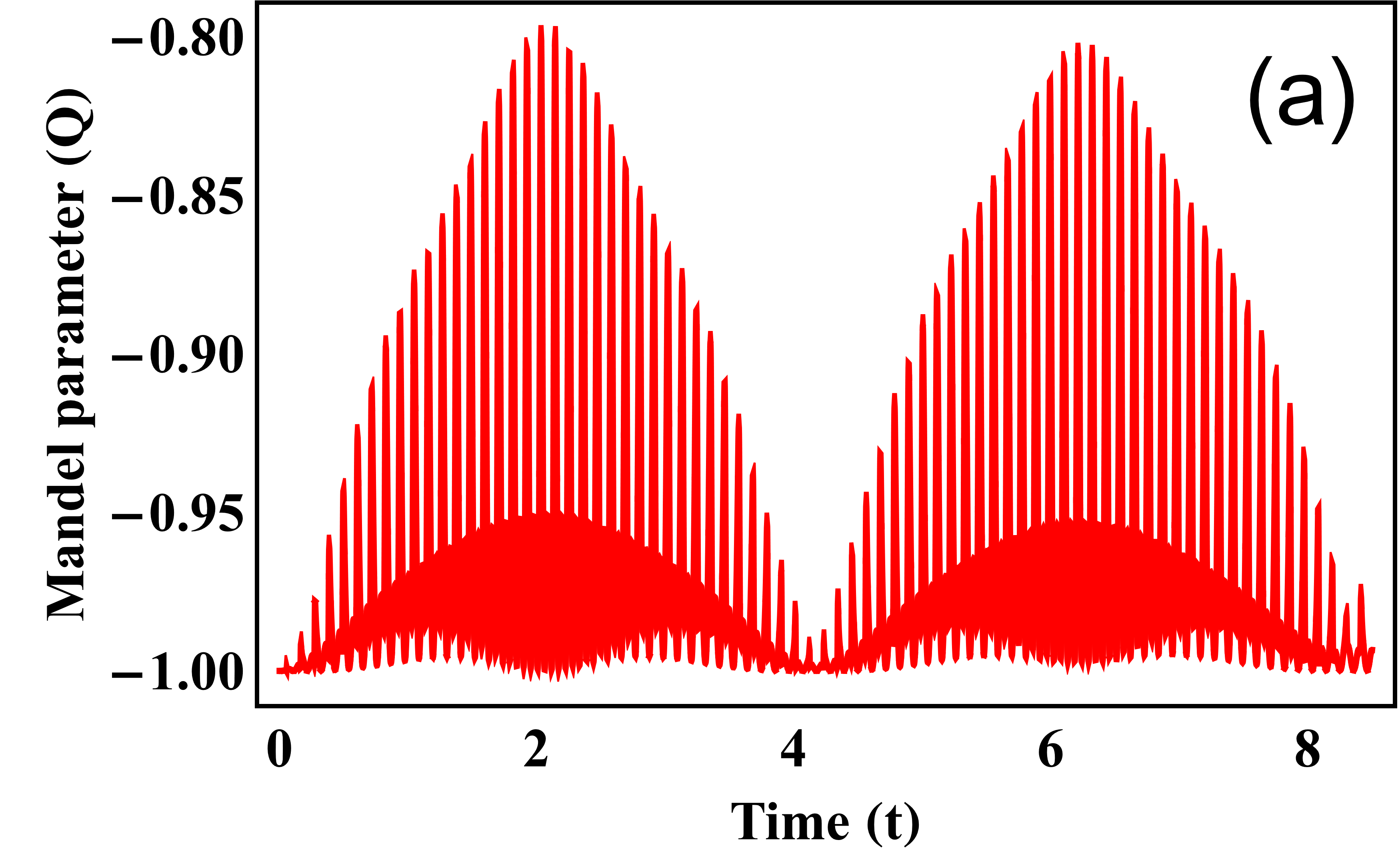}
\includegraphics[scale=0.64]{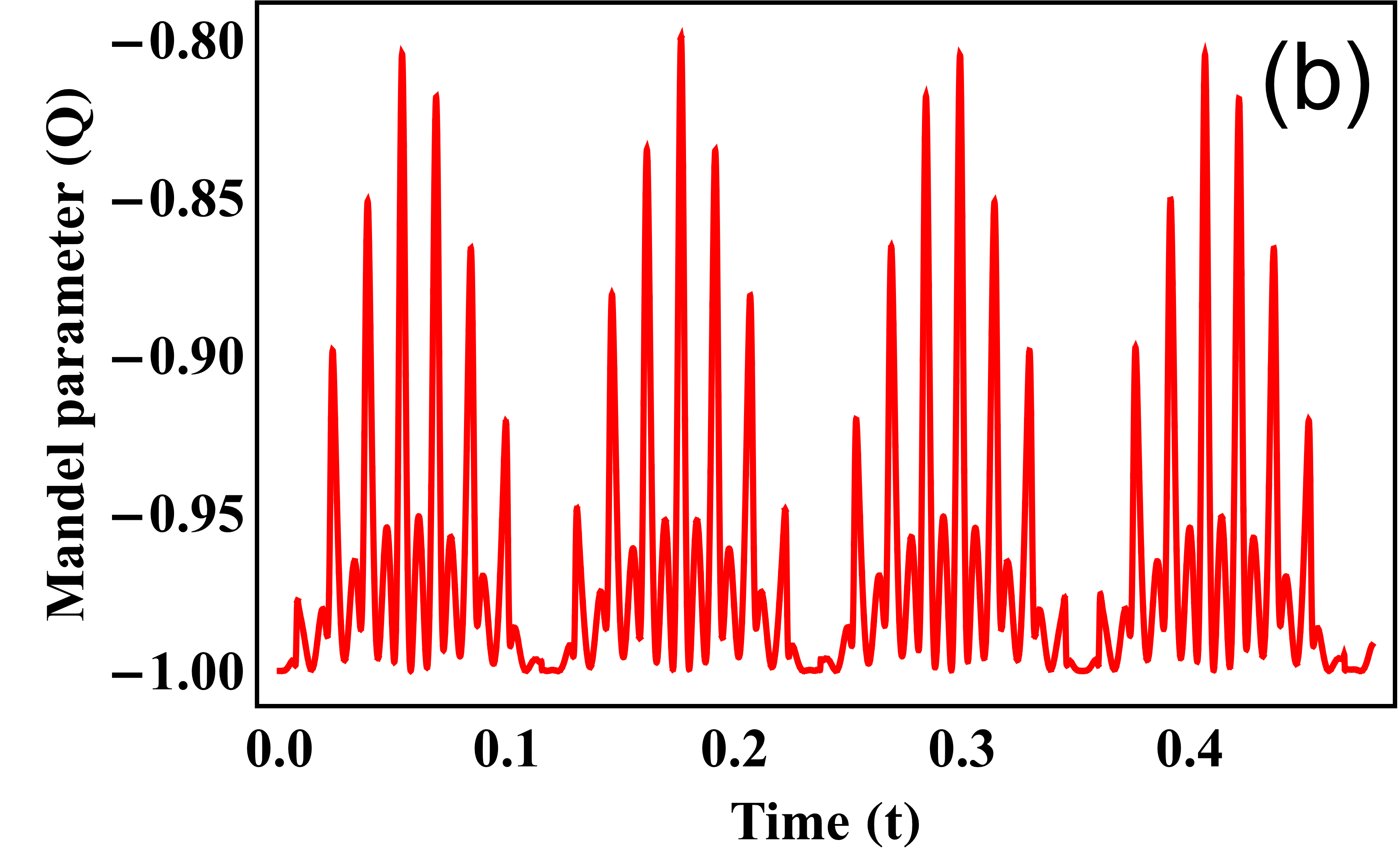}
\caption{Time evolution of the Mandel parameter for $|\alpha|=0.5,\Omega=0.1,\omega=1,N_c=10^5$ (a) $D=0.1$ (b) $D=0.4$}
\label{fig1}
\end{figure}

In this section we shall explore the squeezing properties of the atom laser constructed in the minimal length scenario. First, we shall study the effects analytically and, then, we shall illustrate them numerically. Let us assume that at the beginning the atoms are in the ground state so that the radiation field is initially in a Glauber coherent state $|\psi(0)\rangle_{\text{field}}=|\alpha\rangle_a$, while the initial untrapped excited state is a vacuum state $|\psi(0)\rangle_{\text{atom}}=|0\rangle_b$. Thus, the initial state of the atoms-radiation system is described as $|\psi(0)\rangle=|\alpha\rangle_a\otimes|0\rangle_b$, or alternatively which can also be expressed as $|\psi(0)\rangle=|\alpha/\sqrt{2}\rangle_a\otimes|\alpha/\sqrt{2}\rangle_b$, where $|\alpha/\sqrt{2}\rangle_{a}$ and $|\alpha/\sqrt{2}\rangle_{b}$ are the coherent states associated with the operators $a(t)$ and $b(t)$, respectively. In order to compute the quantum statistical properties, we first require to calculate the expectation value of $a(t)$, which is computed as
\begin{equation}\label{41}
\langle\psi(0)|a(t)|\psi(0)\rangle =\frac{\alpha}{2}e^{-|\alpha|^2}\left(\beta_+e^{-i\delta_+}+\beta_-e^{-i\delta_-}\right),
\end{equation}
where
\begin{alignat}{1}
\beta_{\pm} &= e^{|\alpha|^2\cos\left(\frac{3\alpha_2\mp\alpha_1}{2}t\right)\cos\left(\frac{\alpha_2\pm\alpha_1}{2}t\right)}, \\
\delta_{\pm} &=\left[\frac{\omega}{2}(C+D+1)\pm\alpha_3\right]t+|\alpha|^2 \sin\left(\frac{3\alpha_2\mp\alpha_1}{2}t\right)\cos\left(\frac{\alpha_2\pm\alpha_1}{2}t\right).
\end{alignat}
A similar type of computations yield the other useful expectation values, such as
\begin{alignat}{1}
&\langle a^\dagger(t) a(t)\rangle =\frac{|\alpha|^2}{2}\left[1+e^{-|\alpha|^2\left\{1-\cos(\alpha_1t)\cos(\alpha_2t)\right\}}\cos\left\{2t\alpha_3-|\alpha|^2\sin(\alpha_1t)\cos(\alpha_2t)\right\}\right], \label{44}\\
&\langle a^\dagger(t) a(t)a^\dagger(t) a(t)\rangle =|\alpha|^2\langle a^\dagger(t) a(t)\rangle-\frac{|\alpha|^4}{8}\left[1-e^{-|\alpha|^2\left\{1-\cos(2\alpha_1t)\cos(2\alpha_2t)\right\}}\right. \label{45}\\
&\qquad\qquad\qquad\qquad\qquad\qquad\qquad\qquad\qquad\quad\left.\times\cos\left\{2t\alpha_3-|\alpha|^2\sin(\alpha_1t)\cos(\alpha_2t)\right\}\right].\notag
\end{alignat}
Now, using the result of \myref{41} it is easy to calculate
\begin{equation}\label{46}
\langle a(t)^2\rangle=\frac{\alpha^2}{4}e^{-|\alpha|^2}\left(\gamma_+e^{-i\epsilon_+}+\gamma_-e^{-i\epsilon_-}+\xi e^{-i\lambda}\right),
\end{equation}
where
\begin{alignat}{1}
&\gamma_{\pm} = e^{\frac{|\alpha|^2}{2}\left[\cos\left(4\alpha_2t\right)+\cos\left\{2t(\alpha_2\mp\alpha_1)\right\}\right]}, \\
& \epsilon_{\pm} =\frac{1}{2}\left[4t\left\{\frac{\omega}{2}(C+D+1)\mp\alpha_3\right\}+|\alpha|^2\left\{\sin(4t\alpha_2)+\sin[2t(\alpha_2\mp\alpha_1)]\right\}\right], \\
& \xi=e^{\frac{|\alpha|^2}{2}\left[\cos\left\{2t(\alpha_1-3\alpha_2\right\}+\cos\left\{2t(\alpha_1+3\alpha_2)\right\}\right]}, \\
& \lambda =4t\left\{\frac{\omega}{2}(C+D+1)\mp\alpha_3\right\}-\frac{|\alpha|^2}{2}\left\{\sin[2t(\alpha_1-3\alpha_2)]-\sin[2t(\alpha_1+3\alpha_2)]\right\}.
\end{alignat}
Using \myref{44} and \myref{45}, we can calculate the Mandel parameter \cite{Mandel}
\begin{equation}
Q=\frac{\langle\{a^\dagger(t) a(t)\}^2\rangle-\langle a^\dagger(t) a(t)\rangle^2}{\langle a^\dagger(t) a(t)\rangle}-1,
\end{equation}
which determines the photon statistics of the quantized radiation field. $Q=0$ corresponds to the case of Poissonian distribution, whereas $Q>0$ and $Q<0$ imply to the super-Poissonian and sub-Poissonian cases, respectively. Thus, for a squeezed state of light one should expect the sub-Poissonian $(Q<0)$ nature of the Mandel parameter. Fig. \ref{fig1} shows the numerical behavior of the time evolution of the Mandel parameter for two different values of the deformation parameter. It is evident from the plots that the Mandel parameter is always negative and, thus, indicating the squeezed behavior of the radiation field.

In order to study the quadrature squeezing properties, we first define the quadrature operators $y=\left[a^\dagger(t)+a(t)\right]/\sqrt{2}$  and $z=i\left[a^\dagger(t)-a(t)\right]/\sqrt{2}$, which can be interpreted as the dimensionless position and momentum operators, respectively. It is well-known that for a well-behaved coherent state of light the Heisenberg uncertainty relation is minimized, i.e. $(\Delta y)^2(\Delta z)^2= 1/4$. Whereas, for squeezed light one of the quadratures must be squeezed, so that either $(\Delta y)^2<1/2$ or $(\Delta z)^2<1/2$ should hold. These squeezing conditions can also be expressed as $2(\Delta y)^2-1<0$ and $2(\Delta z)^2-1<0$, so that we can write the relations elegantly for our case as follows
\begin{alignat}{1}
S_1=2(\Delta y)^2-1 &=2\left\{\langle a^\dagger(t) a(t)\rangle +\text{Re}[\langle a^2(t)\rangle]-\text{Re}[\langle a(t)\rangle^2]-|\langle a(t)\rangle |^2\right\}, \label{412}\\
S_2=2(\Delta z)^2-1 &=2\left\{\langle a^\dagger(t) a(t)\rangle -\text{Re}[\langle a^2(t)\rangle]+\text{Re}[\langle a(t)\rangle^2]-|\langle a(t)\rangle |^2\right\}. \label{413}
\end{alignat}
Therefore, the radiation field can be said to be quadrature squeezed if either $S_1$ or $S_2$ is negative. A numerical analysis can be performed easily with the help of the following expressions
\begin{figure}
\centering \includegraphics[scale=0.64]{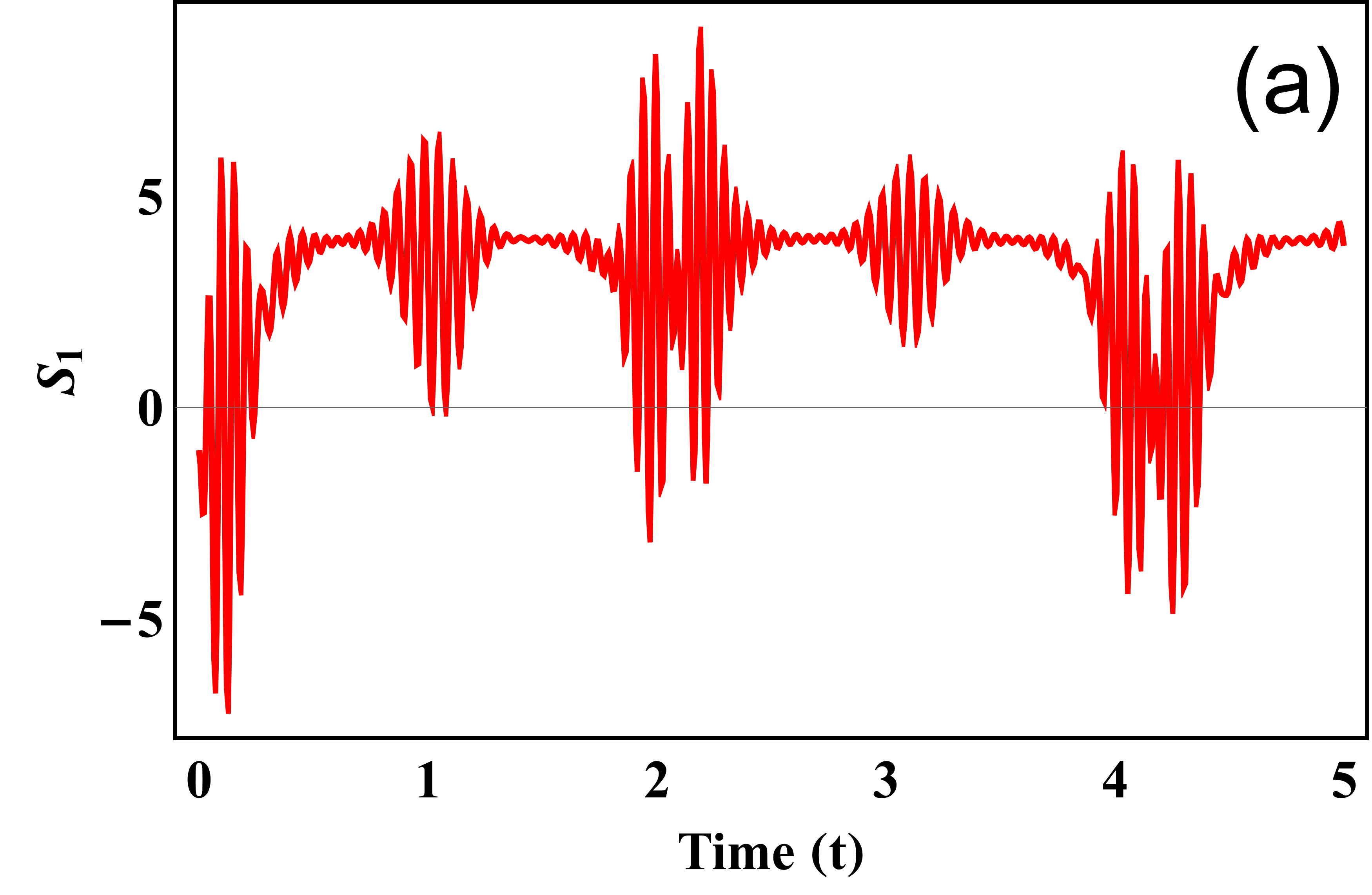}
\includegraphics[scale=0.64]{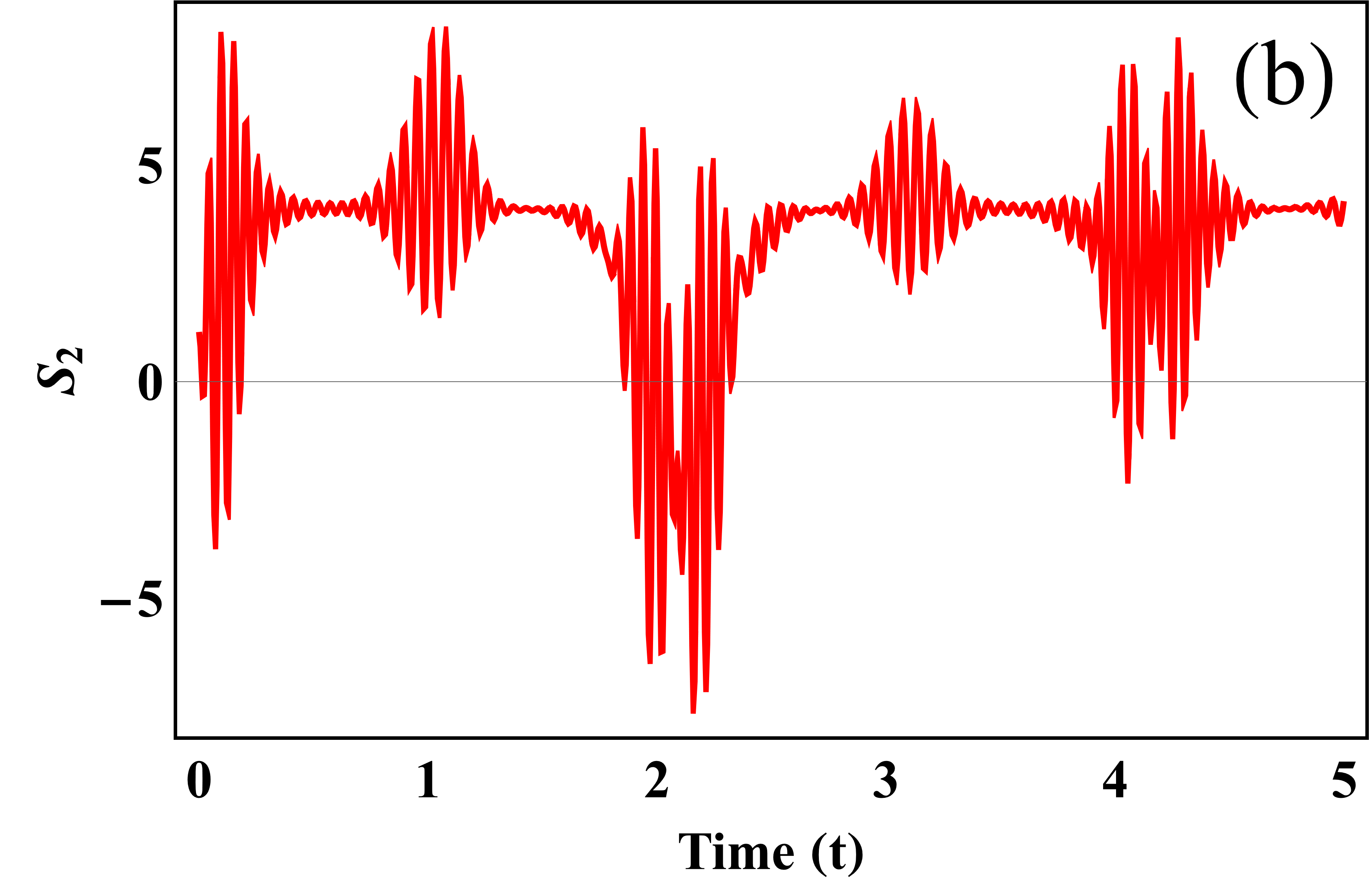}
\caption{Time evolution of (a) $S_1$ (b) $S_2$ for $|\alpha|=2,\Omega=0.2,\omega=1,N_c=10^5, D=0.1, \theta=0.5$. }
\label{fig2}
\end{figure}
\begin{alignat}{1}
&\text{Re}\left[\langle a(t)\rangle^2\right]=\frac{|\alpha|^2}{4}e^{-2|\alpha|^2}\left[2\beta_+\beta_-\cos(\delta_++\delta_--2\theta)+\beta_+^2\cos(2\delta_+-2\theta)+\beta_-^2\cos(2\delta_--2\theta)\right], \label{414}\\
&\text{Re}\left[\langle a^2(t)\rangle\right]=\frac{|\alpha|^2}{4}e^{-|\alpha|^2}\left[\gamma_+\cos(2\theta-\epsilon_+)+\gamma_-\cos(2\theta-\epsilon_-)+\xi\cos(2\theta-\lambda)\right], \label{415}
\end{alignat}
which have been obtained from \myref{41} and \myref{46}, respectively. Here the parameter of the coherent state $\alpha$ is decomposed as $\alpha=|\alpha|e^{i\theta}$. Thus, by replacing \myref{414} and \myref{415} into \myref{412} and \myref{413}  we can study the quadrature squeezing behavior of the radiation field easily, which is demonstrated in Fig. \ref{fig2}. In panel (a), we demonstrate the time-evolution of the squeezing parameter $S_1$ corresponding to the $y$ quadrature, whereas panel (b) shows the behavior of that of $S_2$, which is associated to the $z$ quadrature. We notice that in some regions both of the quadrature become negative and, thus, squeezed. The same phenomena happen for other values of the deformation parameter $D$, which we do not present here.
\section{Concluding remarks}\label{sec5} 
The existence of minimal length is supported by many quantum theories of gravity and, thus, noncommutative theories have become popular as one of the candidates of quantum gravity. However, it is also known by now that the noncommutative theories are not directly compatible with standard quantum mechanics, but with its slight deformed version, without violating any basic laws of quantum mechanics. Therefore, it has become an important area of research to study the deformed quantum mechanical phenomena in different areas of quantum optics and information theory, where the effects of quantum mechanics are easily visible. As it turns out that people do not find entirely new phenomena, but some interesting consequences follow up. For instance, people often end up with systems having more degrees of freedom or control by which one can deal with the system in a much better way. In this paper, we have explored the atom laser for the deformed quantum mechanical structure emerging from the noncommutative framework. The most interesting outcome of this study is to find a squeezed atom laser without the help of any external squeezed light signal as its input. This can not be achieved with the ordinary quantum mechanical atom laser. In order to obtain a squeezed atom laser, people have to take either the nonlinear interaction between the atom and the beam, or one has to send the squeezed light in its input. However, in our system we do not require any of them, but the squeezing properties are inherited by the model by its construction. On top of that, our model is compatible with quantum gravity theories. 

\vspace{0.5cm} \noindent \textbf{\large{Acknowledgements:}} S.D. would like to thank Mir Faizal for useful discussions. S.D. is supported by an INSPIRE Faculty Grant (DST/INSPIRE/04/2016/001391) by the Department of Science and Technology, Govt. of India. V.H. acknowledges the support of research grants from CRSNG of Canada.


\begin{thebibliography}{100}

\bibitem{Aspect_Dalibard_Roger}
A.~Aspect, J.~Dalibard and G.~Roger, Experimental test of Bell's inequalities using time-varying analyzers, Phys. Rev. Lett. \textbf{49}, 1804 (1982).

\bibitem{Ou_Pereira_Kimble_Peng}
Z.~Ou, S.~F. Pereira, H.~J. Kimble and K.~C. Peng, Realization of the Einstein-Podolsky-Rosen paradox for continuous variables, Phys. Rev. Lett. \textbf{68}, 3663 (1992).

\bibitem{Furusawa_et_al}
A.~Furusawa~et al., Unconditional quantum teleportation, Science, \textbf{282}, 706--709 (1998).

\bibitem{Bowen_Lam_Ralph}
W.~P. Bowen, P.~K. Lam and T.~C. Ralph,
\newblock Biased {EPR} entanglement and its application to teleportation,
\newblock {J. Mod. Opt.} \textbf{50}, 801--813 (2003).

\bibitem{Ketterle_Noble}
W.~Ketterle,
\newblock Nobel lecture: {When atoms behave as waves: Bose-Einstein condensation and the atom laser},
\newblock {Rev. Mod. Phys.} \textbf{74}, 1131 (2002).

\bibitem{Kheruntsyan_Olsen_Drummond}
K.~V. Kheruntsyan, M.~K. Olsen and P.~D. Drummond,
\newblock {Einstein-Podolsky-Rosen correlations via dissociation of a molecular Bose-Einstein condensate},
\newblock {Phys. Rev. Lett.} \textbf{95}, 150405 (2005).

\bibitem{Haine_Hope}
S.~A. Haine and J.~J. Hope,
\newblock Outcoupling from a {Bose-Einstein} condensate with squeezed light to produce entangled-atom laser beams,
\newblock {Phys. Rev. A} \textbf{72}, 033601 (2005).

\bibitem{Peters_Chung_Young_Hensley_Chu}
A.~Peters~et al.,
\newblock Precision atom interferometry,
\newblock {Phil. Trans. R. Soc. A} \textbf{355}, 2223 (1997).

\bibitem{Ketterle}
W.~Ketterle,
\newblock Experimental studies of {Bose-Einstein} condensation,
\newblock {Phys. Today} \textbf{52}, 30--35 (1999).

\bibitem{Dowling}
J.~P. Dowling,
\newblock Correlated input-port, matter-wave interferometer: {Quantum-noise} limits to the atom-laser gyroscope,
\newblock {Phys. Rev. A} \textbf{57}, 4736 (1998).

\bibitem{Jing_Chen_Ge}
H.~Jing, J.-L. Chen and M.-L. Ge,
\newblock Quantum-dynamical theory for squeezing the output of a {Bose-Einstein} condensate,
\newblock {Phys. Rev. A} \textbf{63}, 015601 (2000).

\bibitem{Fleischhauer_Gong}
M.~Fleischhauer and S.~Gong,
\newblock Stationary source of nonclassical or entangled atoms,
\newblock {Phys. Rev. Lett.} \textbf{88}, 070404 (2002).

\bibitem{Haine_Olsen_Hope}
S.~A. Haine, M.~K. Olsen and J.~J. Hope,
\newblock Generating controllable atom-light entanglement with a {Raman atom laser system},
\newblock {Phys. Rev. Lett.} \textbf{96}, 133601 (2006).

\bibitem{Tan_Li_Ficek}
R.~Tan, G.-X. Li and Z.~Ficek,
\newblock Squeezed single-atom laser in a photonic crystal,
\newblock {Phys. Rev. A} \textbf{78}, 023833 (2008).

\bibitem{Ries_Brezger_Lvovsky}
J.~Ries, B.~Brezger and A.~I. Lvovsky,
\newblock Experimental vacuum squeezing in rubidium vapor via self-rotation,
\newblock {Phys. Rev. A} \textbf{68}, 025801 (2003).

\bibitem{Tanimura_et_al}
T.~Tanimura~et al.,
\newblock Generation of a squeezed vacuum resonant on a {rubidium D1 line with periodically poled $KTiOPO_4$},
\newblock {Opt. Lett.} \textbf{31}, 2344--2346 (2006).

\bibitem{Hetet_et_al}
G.~H{\'e}tet~et al.,
\newblock Squeezed light for bandwidth-limited atom optics experiments at the rubidium {D1 line},
\newblock {J. Phys. B: Atom. Mol. Opt. Phys.} \textbf{40}, 221 (2006).

\bibitem{Mccormick_Boyer_Arimondo_Lett}
C.~F. McCormick, V.~Boyer, E.~Arimondo and P.~D. Lett,
\newblock Strong relative intensity squeezing by four-wave mixing in rubidium vapor,
\newblock {Opt. Lett.} \textbf{32}, 178--180 (2007).

\bibitem{Jing_Chen_Ge_1}
H.~Jing, J.-L. Chen and M.-L. Ge,
\newblock Squeezing effects of an atom {laser: Beyond the linear model},
\newblock {Phys. Rev. A} \textbf{65}, 015601 (2001).

\bibitem{Johnsson_Haine}
M.~T. Johnsson and S.~A. Haine,
\newblock Generating quadrature squeezing in an atom laser through self-interaction,
\newblock {Phys. Rev. Lett.} \textbf{99}, 010401 (2007).

\bibitem{Dey_Review}
S.~Dey, A.~Fring and V.~Hussin,
\newblock A squeezed review on coherent states and nonclassicality for {non-Hermitian} systems with minimal length,
\newblock {arXiv:1801.01139}.

\bibitem{Dey_Fring_squeezed}
S.~Dey and A.~Fring,
\newblock Squeezed coherent states for noncommutative spaces with minimal length uncertainty relations,
\newblock {Phys. Rev. D} \textbf{86}, 064038 (2012).

\bibitem{Dey_Hussin}
S.~Dey and V.~Hussin,
\newblock Entangled squeezed states in noncommutative spaces with minimal length uncertainty relations,
\newblock {Phys. Rev. D} \textbf{91}, 124017 (2015).

\bibitem{Dey_Hussin_2}
S.~Dey and V.~Hussin,
\newblock Noncommutative $q$-photon-added coherent states,
\newblock {Phys. Rev. A} \textbf{93}, 053824 (2016).

\bibitem{Dey_Fring_Hussin}
S.~Dey, A.~Fring and V.~Hussin,
\newblock Nonclassicality versus entanglement in a noncommutative space,
\newblock {Int. J. Mod. Phys. B} \textbf{31}, 1650248 (2017).

\bibitem{Snyder}
H.~S. Snyder,
\newblock Quantized space-time,
\newblock {Phys. Rev.} \textbf{71}, 38 (1947).

\bibitem{Seiberg_Witten}
N.~Seiberg and E.~Witten,
\newblock String theory and noncommutative geometry,
\newblock {J. High Energy Phys.} \textbf{1999}, 032 (1999).

\bibitem{Connes_Book}
A.~Connes,
\newblock {\em Noncommutative geometry},
\newblock Academic Press (1995).

\bibitem{Garay}
L.~J Garay,
\newblock Quantum gravity and minimum length,
\newblock {Int. J. Mod. Phys. A} \textbf{10}, 145--165 (1995).

\bibitem{Madore_Book}
J.~Madore,
\newblock {\em An introduction to noncommutative differential geometry and its physical applications}.
\newblock Cambridge Univ. Press: UK (1999).

\bibitem{Amelino_Majid}
G.~Amelino-Camelia and S.~Majid,
\newblock Waves on noncommutative space--time and gamma-ray bursts,
\newblock {Int. J. Mod. Phys. A} \textbf{15}, 4301--4323 (2000).

\bibitem{Douglas_Nekrasov}
M.~R. Douglas and N.~A. Nekrasov,
\newblock Noncommutative field theory,
\newblock {Rev. Mod. Phys.} \textbf{73}, 977 (2001).

\bibitem{Szabo}
R.~J. Szabo,
\newblock Quantum field theory on noncommutative spaces,
\newblock {Phys. Rep.} \textbf{378}, 207--299 (2003).

\bibitem{Kempf_Mangano_Mann}
A.~Kempf, G.~Mangano and R.~B. Mann,
\newblock Hilbert space representation of the minimal length uncertainty relation,
\newblock {Phys. Rev. D} \textbf{52}, 1108 (1995).

\bibitem{Das_Vagenas}
S.~Das and E.~C. Vagenas,
\newblock Universality of quantum gravity corrections,
\newblock {Phys. Rev. Lett.} \textbf{101}, 221301 (2008).

\bibitem{Gomes_Kupriyanov}
M.~Gomes and V.~G. Kupriyanov,
\newblock Position-dependent noncommutativity in quantum mechanics,
\newblock {Phys. Rev. D} \textbf{79}, 125011 (2009).

\bibitem{Bagchi_Fring}
B.~Bagchi and A.~Fring,
\newblock Minimal length in quantum mechanics and {non-Hermitian Hamiltonian systems},
\newblock {Phys. Lett. A} \textbf{373}, 4307--4310 (2009).

\bibitem{Quesne_Tkachuk}
C.~Quesne and V.~M. Tkachuk,
\newblock Composite system in deformed space with minimal length,
\newblock {Phys. Rev. A} \textbf{81}, 012106 (2010).

\bibitem{Pedram_Nozari_Taheri}
P.~Pedram, K.~Nozari and S.~H. Taheri,
\newblock The effects of minimal length and maximal momentum on the transition rate of ultra cold neutrons in gravitational field,
\newblock {J. High Energy Phys.} \textbf{2011}, 1--11 (2011).

\bibitem{Mead}
C.~A. Mead,
\newblock Possible connection between gravitation and fundamental length,
\newblock {Phys. Rev.} \textbf{135}, B849 (1964).

\bibitem{Maggiore}
M.~Maggiore,
\newblock A generalized uncertainty principle in quantum gravity,
\newblock {Phys. Lett. B} \textbf{304}, 65--69 (1993).

\bibitem{Scardigli}
F.~Scardigli,
\newblock Generalized uncertainty principle in quantum gravity from micro-black hole gedanken experiment,
\newblock {Phys. Lett. B} \textbf{452}, 39--44 (1999).

\bibitem{Ng_VanDam}
Y.~J. Ng and H.~Van~Dam,
\newblock Spacetime foam, holographic principle, and black hole quantum computers,
\newblock {Int. J. Mod. Phys. A} \textbf{20}, 1328--1335 (2005).

\bibitem{Park}
M.~Park,
\newblock The generalized uncertainty principle {in (A)dS} space and the modification of hawking temperature from the minimal length,
\newblock {Phys. Lett. B} \textbf{659}, 698--702 (2008).

\bibitem{Bhat_Dey_Faizal_Hou_Zhao}
A.~Bhat, S.~Dey, M.~Faizal, C.~Hou and Q.~Zhao,
\newblock Modification of {Schr{\"o}dinger--Newton} equation due to braneworld models with minimal length,
\newblock {Phys. Lett. B} \textbf{770}, 325--330 (2017).

\bibitem{Rovelli}
C.~Rovelli,
\newblock Loop quantum gravity,
\newblock {Living Rev. Rel.} \textbf{1}, 1 (1998).

\bibitem{Dzierzak}
P.~Dzierzak, J.~Jezierski, P.~Malkiewicz and W.~Piechocki,
\newblock The minimum length problem of loop quantum cosmology,
\newblock {Acta Phys. Pol. B} \textbf{41}, 717--726 (2010).

\bibitem{Veneziano}
G.~Veneziano,
\newblock A stringy nature needs just two constants,
\newblock {Europhys. Lett.} \textbf{2}, 199 (1986).

\bibitem{Gross_Mende}
D.~J. Gross and P.~F. Mende,
\newblock String theory beyond the Planck scale,
\newblock {Nucl. Phys. B} \textbf{303}, 407--454 (1988).

\bibitem{Amati}
D.~Amati, M.~Ciafaloni and G.~Veneziano,
\newblock Can spacetime be probed below the string size?
\newblock {Phys. Lett. B} \textbf{216}, 41--47 (1989).

\bibitem{Yoneya}
T.~Yoneya,
\newblock On the interpretation of minimal length in string theories,
\newblock {Mod. Phys. Lett. A} \textbf{4}, 1587--1595 (1989).

\bibitem{Konishi}
K.~Konishi, G.~Paffuti and P.~Provero,
\newblock Minimum physical length and the generalized uncertainty principle in string theory,
\newblock {Phys. Lett. B} \textbf{234}, 276--284 (1990).

\bibitem{Greensite}
J.~Greensite,
\newblock Is there a minimum length in $d=4$ lattice quantum gravity?
\newblock {Phys. Lett. B} \textbf{255}, 375--380 (1991).

\bibitem{Padmanabhan}
T.~Padmanabhan,
\newblock Physical significance of {Planck} length,
\newblock {Ann. Phys.} \textbf{165}, 38--58 (1985).

\bibitem{Amelino}
G.~Amelino-Camelia,
\newblock Testable scenario for relativity with minimum length,
\newblock {Phys. Lett. B} \textbf{510}, 255--263 (2001).

\bibitem{Magueijo}
J.~Magueijo and L.~Smolin,
\newblock Lorentz invariance with an invariant energy scale,
\newblock {Phys. Rev. Lett.} \textbf{88}, 190403 (2002).

\bibitem{Pikovski}
I.~Pikovski, M.~R. Vanner, M.~Aspelmeyer, M.~Kim and C~Brukner,
\newblock {Nature Phys.} \textbf{8}, 393--397 (2012).

\bibitem{Dey_NPB}
S.~Dey~et al.,
\newblock Probing noncommutative theories with quantum optical experiments,
\newblock {Nucl. Phys. B} \textbf{924}, 578--587 (2017).

\bibitem{Dey_SciRep}
M.~Khodadi, K.~Nozari, S.~Dey, A.~Bhat and M.~Faizal,
\newblock A new bound on polymer quantization via an opto-mechanical setup,
\newblock {Sci. Rep.} \textbf{8}, 1659 (2018).

\bibitem{Ng_Review}
Y.~J. Ng,
\newblock Selected topics in Planck-scale physics,
\newblock {Mod. Phys. Lett. A} \textbf{18}, 1073--1097 (2003).

\bibitem{Hossenfelder_Review}
S.~Hossenfelder,
\newblock Minimal length scale scenarios for quantum gravity,
\newblock {Living Rev. Relat.} \textbf{16}, 2 (2013).

\bibitem{Dey_Fring_Gouba}
S.~Dey, A.~Fring and L.~Gouba,
\newblock $\mathcal{PT}$-symmetric non-commutative spaces with minimal volume uncertainty relations,
\newblock {J. Phys. A: Math. Theor.} \textbf{45}, 385302 (2012).

\bibitem{Bender_Boettcher}
C.~M. Bender and S.~Boettcher,
\newblock Real spectra in {non-Hermitian Hamiltonians having $\mathcal{PT}$-symmetry},
\newblock {Phys. Rev. Lett.} \textbf{80}, 5243 (1998).

\bibitem{Bender_Making_Sense}
C.~M. Bender,
\newblock Making sense of {non-Hermitian Hamiltonians},
\newblock {Rept. Prog. Phys.} \textbf{70}, 947 (2007).

\bibitem{Scholtz_Geyer_Hahne}
F.~G. Scholtz, H.~B. Geyer and F.~Hahne,
\newblock {Quasi-Hermitian} operators in quantum mechanics and the variational principle,
\newblock {Ann. Phys.} \textbf{213}, 74--101 (1992).

\bibitem{mostafazadeh1}
A.~Mostafazadeh,
\newblock {Pseudo-Hermiticity versus $\mathcal{PT}$ symmetry: the necessary condition for the reality of the spectrum of a non-Hermitian Hamiltonian},
\newblock {J. Math. Phys.} \textbf{43}, 205 (2002).

\bibitem{Mewes}
M.-O. Mewes~et al.,
\newblock Output coupler for {Bose-Einstein} condensed atoms,
\newblock {Phys. Rev. Lett.} \textbf{78}, 582 (1997).

\bibitem{Changpu}
S.~Changpu, Z.~He, M.~Yuanxiu and L.~Jianming,
\newblock On the quantum dynamic theory of the {MIT output coupler for the Bose-Einstein} condensation,
\newblock {Commun. Theor. Phys.} \textbf{29}, 161 (1998).

\bibitem{Bogoliubov}
N.~Bogoliubov,
\newblock On the theory of superfluidity,
\newblock {J. Phys. (USSR)} \textbf{11}, 23 (1947).

\bibitem{Kuang_Ouyang}
L.-M. Kuang and Z.-W. Ouyang,
\newblock Macroscopic quantum self-trapping and atomic tunneling in two-species {Bose-Einstein} condensates,
\newblock {Phys. Rev. A} \textbf{61}, 023604 (2000).

\bibitem{Mandel}
L.~Mandel,
\newblock {Sub-Poissonian} photon statistics in resonance fluorescence,
\newblock {Opt. Lett.} \textbf{4}, 205--207 (1979).

\end{thebibliography}


\end{document}